\documentclass{article}
\usepackage{amssymb}
\usepackage{amsmath}

\input{tcilatex}

\begin{document}

\title{Nature of the Quantum Potential}
\author{Anton A. Lipovka \\
Centro de Investigacion en Fisica, Universidad de Sonora, \\
Sonora, Mexico. e-mail: aal@cifus.uson.mx}
\date{March 01, 2016 }
\maketitle

\begin{abstract}
In this paper we suggest a natural interpretation of the de Broglie-Bohm
quantum potential, as the energy due to the oscillating electromagnetic
field (virtual photon) coupled with moving charged particle. Generalization
of the Schr\"{o}dinger equation is obtained. The wave function is shown to
be the eigenfunction of the Sturm - Liouville problem in which we expand
virtual photon to include it implicitly into consideration. It is shown the
non - locality of quantum mechanics is related only with virtual photon. As
an example, the zero - energy of harmonic oscillator is obtained from
classical equations.
\end{abstract}

PACS: 02.40, 03.65.-w, 03.65.Ta, 05.30.-d, 11.10.Ef \newline

"I think I can safely say that nobody understands quantum mechanics."

(Richard Feynman)

\bigskip

\section{Introduction}

\bigskip

Interpretation of quantum theory (both classical and relativistic) since its
birth to the present day, is the subject of much debate on the fundaments of
quantum physics. At first glance, the situation is reassuring, because now
we have dozen different formulations of quantum theory [1] and can apply
these powerful techniques for computing. However, there are a number of
unresolved fundamental issues shows us something we do misunderstand in the
meaning of these formulations and we still no have a final interpretation of
quantum theory.

It is well known, the Bohmian formulation based on Louis de Broglie's
pilot-wave theory, suggest particularly a great conceptual advantages in
possible interpretation because it is causal and not local. Finally it leads
to the complete Hamilton function that contains the quantum potential, which
is of great importance and reflects fundamental properties of nature (see
[2,3,4,5] and references therein).

The quantum potential plays a central role in the formalism of Bohm and is
used in modern physics and theoretical chemistry. Actually it is of great
importance because on the one hand the Bohmian formulation and quantum
potential allow us deeply understand the basics of the theory. On the other
hand quantum potential has multiples practical applications in different
fields of knowledge (for example in the solid state physics, in theoretical
chemistry etc.) because it gives us an opportunity to simulate different
quantum effects without the involvement of the wave functions of system, and
without solving the Schr\"{o}dinger equation. In this case, the Monte Carlo
method is applied to the hydrodynamic calculations, which are sometimes the
only possibility to get the result, when the Schr\"{o}dinger equation can
not be solved exactly.

Unfortunately up to now we did not understand clearly the nature of quantum
potential and wave functions. This problem on the one hand provokes many
speculations and strange hypothesis, but on the other hand this
misunderstanding makes it difficult to progress in important areas such as
quantum computers, solid state and many others. These fundamental questions
were debated by many authors from the beginning of quantum theory. As an
example we quote here the paper [1]: \textquotedblleft Schr\"{o}dinger
invented this formulation in hope of casting quantum mechanics into a
\textquotedblleft congenial\textquotedblright\ and \textquotedblleft
intuitive\textquotedblright\ form -- he was ultimately distressed when he
found that his wavefunctions were functions in configuration space and did
not actually exist out in ordinary three-dimensional
space.\textquotedblright ,\ and also: \textquotedblleft The wavefunction
should be regarded as a mathematical tool for calculating the outcomes of
observations, not as a physically present entity existing in space\ldots
\textquotedblright .

As it was mentioned above, this misunderstanding provokes many strange and
exotic ideas and papers, where quantum potential is used for speculations,
particularly in cosmology to construct the most intrigue entity --
cosmological constant. For this reasons it is of great importance to reveal
physical sense and nature of quantum potential, and to determine the
specific place of the wave function in the formalism of quantum mechanics.

Recently the paper [6] was published, in which a new foundation has been
proposed to unify the quantum theory and relativity. In this paper it was
shown that quantization naturally appears as consequence of geometrical
properties of our Universe, and Planck's constant is just adiabatic
invariant of the electromagnetic field, determined by the Riemann - Cartan
geometry (by the Hubble constant and the cosmological constant). It was
constructed self-consistent non-local quantum theory based on Einstein's
generalized theory of gravity in the space of Riemann - Cartan. It should be
stressed this theory does not requires any initial assumptions, external to
the theory and alien by nature. So, we need not any axioms, wave functions
or hidden variables.

Another paper in which a natural origin of quantization is suggested, was
published by Garcia - Morales [7]. In this work the quantization of the
action $S$ was obtained as a consequence of thermodynamic theory (i.e. also
without artificial axiomatic constructions).

These results also suggest that the orthodox formulations of quantum theory
[1] based on the axiom of the wave function existence are not complete and
should be reconsidered.

In present paper the quantum potential is shown to be formed by bounded
electromagnetic field - virtual photon (see section 3 for details), which is
a principle part and main participant for any bounded quantum system.

\section{Enigma of quantum potential.}

\bigskip

Usually quantum potential in the Bohmian formulation of quantum theory is
defined this way (we consider here one - particle case because that for many
particles is treated the same manner). Schr\"{o}dinger equation is

\begin{equation}
i\hbar \frac{\partial \Psi (x,t)}{\partial t}=-\frac{\hbar ^{\mathbf{2}}}{2m}%
\nabla ^{2}\Psi (x,t)+U(x)\Psi (x,t)  \tag{1}
\end{equation}

Writing the wave function in form \bigskip $\Psi (x,t)=R(x,t)\exp
(iS(x,t)/\hbar )$ we immediately obtain two equations:

\begin{equation}
\frac{\partial S}{\partial t}=-H_{tot}=-\frac{(\nabla S)^{2}}{2m}-U(x)+\frac{%
\hbar ^{\mathbf{2}}}{2m}\frac{\nabla ^{2}R}{R}  \tag{2}
\end{equation}

and

\begin{equation}
\frac{\partial R^{2}}{\partial t}+\nabla (R^{2}\frac{\nabla S}{m})=0  \tag{3}
\end{equation}

The first one is a Hamilton - Jacobi equation written for a modified
Hamiltonian:

\begin{equation}
H_{tot}(t,x,p,R)=\frac{(\nabla S)^{2}}{2m}+U(x)-Q(x,t)  \tag{4}
\end{equation}

where $Q(x,t)=\frac{\hbar ^{\mathbf{2}}}{2m}\frac{\nabla ^{2}R}{R}$ is
so-called quantum potential, and second one is a continuity equation written
for $R^{2}$. This modified Hamiltonian usually named as $\Psi $ - dependent
one, and additional term \ (the quantum potential) usually interpreted as
"an internal energy associated with a certain region of phase space, absent
in classical mechanics, but arising in quantum mechanics from the
uncertainty principle" [2]. It is difficult to agree with this point of view
for many reasons. First of all note that the discussed system, described by
equation (1) is supposed to be isolated one, but it does not contains any
variable electromagnetic field. Instead, it contains an artificial quantum
potential.

It is clear, our system should contain an oscillating electromagnetic field
produced by electron, but we do not see it in equation (1). At the same time
the Bohmian formulation has the following features:

- there is presence of hidden variables ( it should be treated as a hint for
presence of bounded electromagnetic field)

- it is causal (so, it should be a classical field theory)

- not local (presence of an electromagnetic field in theory).

Taking into account the fact that the only real substance we have in our
arsenal to capture experimental events - is electromagnetic field (this is
an experimental fact), we can conclude that this field we lost in the
beginning. This "lost" field was found in [6] and now we are able to
identify quantum potential with that bounded non-local classical
electromagnetic field mentioned above.

\section{Nature does not have hidden variables.}

\bigskip

Recently [8] the idea was suggested that quantum potential appears due to a
\textquotedblleft concealed\textquotedblright\ motion. Namely author show
that in the case when our system consist of two parts, one of which is
characterized by observable coordinates $q_{i}(t)$ and other one has
\textquotedblleft concealed\textquotedblright\ coordinates $Q_{i}(t)$ (here
we use notations of author), the Hamilton function of such system can be
shown to contain quantum potential, which \textquotedblleft may be regarded
as the kinetic energy of additional concealed freedoms.\textquotedblright
\lbrack 8]. But the Lagrangian used by author is incomplete (see eq. (1.1)
in [8]), so to begin with we need to write correct action for our system.

Action of the system under consideration one can write as:

\begin{equation}
S_{tot}=S_{0}+S_{int}+S_{em}\text{ \ .}  \tag{5}
\end{equation}

Here $S_{0}$ is action for particle(s) without electromagnetic field,

\begin{equation}
S_{int}=-\frac{1}{c^{2}}\int j^{\alpha }(x)A_{\alpha }(x)d^{4}x  \tag{6}
\end{equation}

- is the interaction between electromagnetic field and charge(s), and

\begin{equation}
S_{em}=-\frac{1}{16\pi c}\int F^{\alpha \beta }F_{\alpha \beta }d^{4}x 
\tag{7}
\end{equation}

- is action for electromagnetic field. It can be seen that both $S_{int}$
and $S_{em}$ contains the field coordinates, so in this case we are speaking
about a bounded photon coupled with electron. Such bounded photon is named
as a virtual photon (see for details [9]) and in this paper this definition
is used.

It should be stressed here, the Bohmian formulation is non-relativistic one
by origin, so we may restrict our speculation by non-relativistic case. Let
us consider the hydrogen atom as an example, where to simplicity sake we
believe $m_{e}<<m_{p}$ . So, we can consider one-body equation, as it was
made in the case of the equation (1). It is easy to show that introduction
of second body will not change main properties of our result, but leads to
small corrections which have well-understood nature, but are not significant
for our present aim.

The classical energy equation for our reduced classical system $S_{0}$
(fixed proton and moving electron) is $H=p^{2}/2m+U(x)=E$ . It is important
emphasize again: in this equation there are no harmonic electromagnetic
field, which must take place due to electron oscillations. Canonical
quantization procedure applied to this relation leads directly to the Schr%
\"{o}dinger equation (1) for a particle moving in a potential $U(x)$.
However the oscillating electron does produce harmonic electromagnetic field 
$\varphi (k,x)$\ (which appears in (6), (7)) and this harmonic function can
be used to write Fourier - transformation of the energy equation on 4 -
coordinates $x$ . This is the way in which electromagnetic field appears
implicitly in the Schr\"{o}dinger equation (1) and Hamilton function
transforms to the operator (see also [6]):%
\begin{equation}
\int \hat{H}\varphi d^{4}x=i\hbar \int \frac{\partial }{\partial t}\varphi
d^{4}x  \tag{8}
\end{equation}

Here integration is carried out over 4-volume , and $\hat{H}$ is operator of
Liouville written for reduced (incomplete) non-relativistic system $\hat{H}%
=-\hbar ^{2}\nabla ^{2}/2m+U(x)$, which corresponds to the problem of Sturm
- Liouville with eigenfunctions $\Psi _{n}(x,t)=\exp (iS_{n}/\hbar )$. These
eigen functions, in turn, form complete basis and we can expand our virtual
photon, and this way include it into consideration through the coefficients
of expansion $R_{n}(p_{\alpha })$:

\begin{equation}
\varphi (p_{\alpha },x^{\alpha })=\dsum R_{n}(p_{\alpha })\Psi
_{n}(x^{\alpha })  \tag{9}
\end{equation}

Here we sum only over $n$ , and index $\alpha $\ is just to mention the fact
we are working with 4-vectors $p_{\alpha }$ and $x^{\alpha }$ in the
Minkowsky space. By substitute this in (8) we have

\begin{equation}
\int \hat{H}\tsum R_{n}(p_{\alpha })\Psi _{n}(x^{\alpha })d^{4}x=i\hbar \int 
\frac{\partial }{\partial t}\tsum R_{n}(p_{\alpha })\Psi _{n}(x^{\alpha
})d^{4}x  \tag{10}
\end{equation}

These expressions actually are complete "quantum" non-local equations to
describe our system in Minkowsky space, with clearly written non-local
hidden variables (beables) of the virtual photon (which are the coefficients 
$R_{n}(p_{\alpha })$). In general case we should integrate this equations
over 4-volume. However if we are interested in non-relativistic Schr\"{o}%
dinger equation, we can evaluate these integrals by taking into account
relation $v_{e}<<c$,\ so, the main part of each integral is contributed by
small region in vicinity of the electron and integration can be carried out
easily. If we are interested in the case when our system stay in a defined
state $n$, we can write (10) as:

\begin{equation}
\left[ -\frac{\hbar ^{\mathbf{2}}}{2m}\nabla ^{2}R-\frac{i\hbar }{m}\nabla
R\nabla S+\frac{R(\nabla S)^{2}}{2m}-\frac{i\hbar }{2m}R\nabla ^{2}S+RU(x)%
\right] =i\hbar \frac{\partial R}{\partial t}-R\frac{\partial S}{\partial t}
\tag{11}
\end{equation}

From this relation we immediately obtain the Hamilton - Jacobi equation (2)
written for a complete Hamiltonian of our system (4), where "quantum
potential" now has clear sense and must be attributed to presence of the
virtual photon (electromagnetic "pilot-wave"). As to the continuity equation
(3) written for $R^{2}$, it must be interpreted as continuity equation for
the square coefficients (of expansion of photon into the eigen functions of
corresponding problem of Sturm -- Liouville) of the expansion of the virtual
photon coupled with electron and moving with velocity of the electron $%
\overset{\rightarrow }{v}_{e}$. It is actually an analog for well known
Poynting's theorem. It should be stressed here in this continuity equation
the Planck constant in fact does not appears, because it is classical
equation for coefficients of expansion \ of classical electromagnetic field
coupled with electron:%
\begin{equation}
\frac{\partial \rho }{\partial t}+\nabla (J)=0  \tag{12}
\end{equation}

where 
\begin{equation}
\rho =R^{2}\text{ ,}  \tag{13}
\end{equation}%
\begin{equation}
\overset{\rightarrow }{J}=R^{2}\frac{\nabla S}{m}=R^{2}\overset{\rightarrow }%
{v}_{e}  \tag{14}
\end{equation}

To conclude this part it should be useful to make some comments on the
Hamilton - Jacobi equation (2)

\begin{equation}
\frac{\partial S}{\partial t}=-H_{tot}=-\frac{(\nabla S)^{2}}{2m}-U(x)+\frac{%
\hbar ^{\mathbf{2}}}{2m}\frac{\nabla ^{2}R}{R}  \tag{15}
\end{equation}

Now, when the physical sense of "quantum potential" (as classical potential
that corresponds to the oscillating electromagnetic field with energy $\hbar
\omega $) became clear, we may definitely interpret limit $\hbar \omega
\rightarrow 0$\ as an hypothetical situation with absence of the virtual
photon (the energy of the oscillating electromagnetic field $\varphi (k,x)$\
is zero). In this case we obtain a classical system with classical Hamilton
function $H=\frac{(\nabla S)^{2}}{2m}+U(x)$ for our incomplete system.

\section{Discussion}

\bigskip

Very fundamental and at the same time useful example suggests harmonic
oscillator. We have discussed it from this point of view before in [6], but
it would be interesting to consider it briefly in respect to the quantum
potential of our system. It should be done the same way as it was made in
previous part (see eq. (5) - (15)). The Hamiltonian written for harmonic
oscillator is:

\begin{equation}
H=-\frac{\hbar ^{\mathbf{2}}}{2m}\nabla ^{2}+\frac{1}{2}m\omega ^{2}r^{2} 
\tag{16}
\end{equation}

Substituting ground state wave function $\Psi =\exp (-m\omega r^{2}/2\hbar )$
of corresponding problem of Sturm -- Liouville into equation (10),
immediately give us quantum potential for harmonic oscillator:

\begin{equation}
Q=\hbar \omega /2  \tag{17}
\end{equation}

which should be recognized as energy of virtual photon in zero-state of
harmonic oscillator (remember here the frequency of electron oscillations is
the same that has the virtual photon). So total Hamilton function for
"quantum" (in reality classical) harmonic oscillator in ground state is

\begin{equation}
H_{tot}(t,x,p,R)=\frac{(\nabla S)^{2}}{2m}+U(x)-\hbar \omega /2  \tag{18}
\end{equation}

with the oscillation frequency of electron (and virtual photon) $\omega $.
It should be noted here, in the paper [2] there is an error in the
expression for the quantum potential of harmonic oscillator, which we have
corrected in eq. (17) and (18).

One can see again - the total Hamilton function corresponds to the complete
mechanical system classical by nature which does not contain the hidden
variables. And so called "quantum potential" appears due to the virtual
photon with frequency $\omega $, which forms a "quantum" zero-state energy $%
\hbar \omega /2$. So, the quantum potential and hidden variables can be
naturally identified with oscillating electromagnetic field (virtual photon)
coupled with the moving electron.

In light of these results it becomes immediately obvious meaning of Bell's
theorem, as a classical statement about the impossibility of motions with a
speed faster than light in the framework of the relativistic theory.

Conclusions of our work could be formulated as follows:

The quantum potential is shown to be an additional energy, electromagnetic
by origin, which appears due to coupled harmonic electromagnetic field
(virtual photon).

The wave functions are shown to be just a complete basis of the Sturm -
Liouville problem (written for reduced system with action $S_{0}$), in which
the virtual photon is expanded to include it implicitly into Schr\"{o}dinger
equation.

It is stressed - the non-locality of quantum mechanics is related only with
this virtual photon, namely with distribution of harmonic electromagnetic
field in the system under consideration.

\ \ \ \ \ \ \ \ \ \ \ \ \ \ \ \ \ \ \ \ \ \ \ \ \ \ \ \ \ \ \ \ \ \ \ \ \ \
\ \ \ \ \ \ \ \ \ \ \ \ \ \ \ \ \ \ \ \ \ \ \ \ \ \ \ \ \ \ \ \ \ 

\section{Bibliography}

\bigskip

[1] Daniel F. Styer, Miranda S. Balkin, Kathryn M. Becker, Matthew R. Burns,
Christopher E. Dudley, Scott T. Forth, Jeremy S. Gaumer, Mark A. Kramer,
David C. Oertel, Leonard H. Park, Marie T. Rinkoski, Clait T. Smith, and
Timothy D. Wotherspoon. (March 2002) Nine formulations of quantum mechanics.
Am. J. Phys. 70 (3) . DOI: 10.1119/1.1445404

[2] len Dennis, aurice A. de Gosson, asil J. Hiley. (June 2015) Bohm's
Quantum Potential as an Internal Energy. Physics Letters A., Volume 379,
Issues 18--19, 26 , pp. 1224--1227. arXiv:1412.5133v1 [quant-ph] 15 Dec 2014

[3] Peter J. Riggs. (2008) Reflections on the deBroglie--Bohm Quantum
Potential.\ Erkenn 68:21--39. DOI: 10.1007/s10670-007-9054-1

[4] S. Esposito. (1999) Photon Wave Mechanics: A De Broglie-Bohm. Approach.
Foundations of Physics Letters, Vol. 12, No. 6, pp.533-545.

[5] Gerhard Gr\"{o}ssing. (2009) On the Thermodynamic Origin of the Quantum
Potential.\ \ Physica A 388 pp. 811-823. \ DOI: 10.1016/j.physa.2008.11.033

[6] Lipovka, A. (2014) Planck Constant as Adiabatic Invariant Characterized
by Hubble's and Cosmological Constants. Journal of Applied Mathematics and
Physics, 2, 61-71. doi: 10.4236/jamp.2014.25009. \ ;
http://lanl.arxiv.org/abs/1401.2404

[7] Vladimir Garcia-Morales. (2015) Quantum Mechanics and the Principle of
Least Radix Economy. Foundations of Physics 45, 295-332 (Springer)

[8] Peter Holland. (2015) Quantum potential energy as concealed motion.
Foundations of Physics (Springer), V. 45, Issue 2, pp 134-141

[9] V. Ginzburg (1987) Theoretical Physics and Astrophysics.\ (Pergamon
press)

\end{document}